\documentclass[prl,twocolumn,floatix,amssymb,showpacs,amsmath,superscriptaddress,usenames,dvipsnames]{revtex4-1}

\usepackage[T1]{fontenc}
\usepackage[dvipsnames]{xcolor}
\usepackage[colorlinks=true,
            linkcolor=RedViolet,
            urlcolor=Magenta,
            citecolor=Blue,
            anchorcolor=red]{hyperref}
\usepackage{amsmath, dsfont, bm}
\usepackage{physics,braket}
\usepackage{graphicx}
\usepackage{mathptmx}
\usepackage[normalem]{ulem}
\usepackage[makeroom]{cancel}

\begin{document}
\title{Robust Macroscopic Matter-Wave Interferometry with Solids}
\author{Julen S. Pedernales and Martin B. Plenio}
\affiliation{Institut f\"ur Theoretische Physik und IQST, Albert-Einstein-Allee 11, Universit\"at Ulm, D-89081 Ulm, Germany}

\begin{abstract}
Matter-wave interferometry with solids is highly susceptible to minute fluctuations of environmental fields, including gravitational effects from distant sources. Hence, experiments require a degree of shielding that is extraordinarily challenging to achieve in realistic terrestrial or even space-based set-ups. Here, we design protocols that exploit the spatial correlations that are inherent in perturbations due to distant sources to reduce significantly their impact on the visibility of interference patterns. We show that interference patterns that are robust to such type of noise can be encoded in the joint probability distribution of two or more interferometers, provided that these are initialized in suitable states. We develop a general framework that makes use of  $N+1$ interferometers that may differ in their masses to correct for environmental potential fields up to order $N$ in their multipole expansion. Remarkably, our approach works for fields that fluctuate stochastically in any time scale and does not require the presence of quantum correlations among the different interferometers. Finally, we also show that the same ideas can be extended to the protection of entanglement between pairs of interferometers.
\end{abstract}

\maketitle

{\it Introduction.---}Performing interferometry with objects of increasing mass has been a constant aspiration of experiment and a source of insight into the laws of nature, ever since de Broglie first suggested the wave character of matter in the early years of quantum mechanics. To date interference has been achieved for a plethora of massive systems, including electrons~\cite{Hasselbach2009}, neutrons~\cite{RW2000}, BECs~\cite{Cornell1998}, fullerenes~\cite{Zeilinger1999}, or macromolecules with up to 25,000~Da~\cite{Tuxen2013,Arndt2019}. Adding to this catalog interferometric experiments performed with solids is, arguably, one of the most ambitious and exciting missions of current quantum technologies. With the advent of optomechanics~\cite{AspelmeyerKM14,TeufelDL2011,ChanMS2011,RiedingerWM2018,RiedingerWM2018,OckeloenDP2018}, and in particular recent results in levitated optomechanics~\cite{MillenMP+19,DRD+20}, the notion that the center of mass (c.m.) of a single solid could be placed in a superposition and made to exhibit interference phenomena~\cite{Cirac2011, Bose2013, Duan2013,Albrecht14, Plenio2020} is perhaps more realistic than ever before.

The motivation to push matter-wave interferometry towards higher and higher masses is twofold. First, from a fundamental point of view, the observation of an interference pattern in the probability distribution of the mechanical degrees of freedom (d.f.) of a massive object tells us that the object was indeed in a coherent superposition. When this superposition concerns large spatial separations, such an experiment becomes a fundamental test of the linearity of quantum mechanics at larger scales~\cite{Tombesi1997,Knight1999,Bouwmeester2003,Hornberger2014}, and could serve as a test bed for alternative descriptions of nature such as those advocating collapse models~\cite{Romero-Isart2011, Ulbricht2013}. Second, the extreme sensitivity of interferometers to external fields that are inhomogeneous across the different interferometric paths makes of large mass matter-wave interferometers exquisite metrological devices of minute forces, such as those of gravitational origin~\cite{Schmoele16,Bose2018, Li2020,WestphalHP+21,Plenio2020+,Plenio2021,Plenio2021+}. 

\begin{figure}[htbp]
   \centering
   \includegraphics[width=\columnwidth]{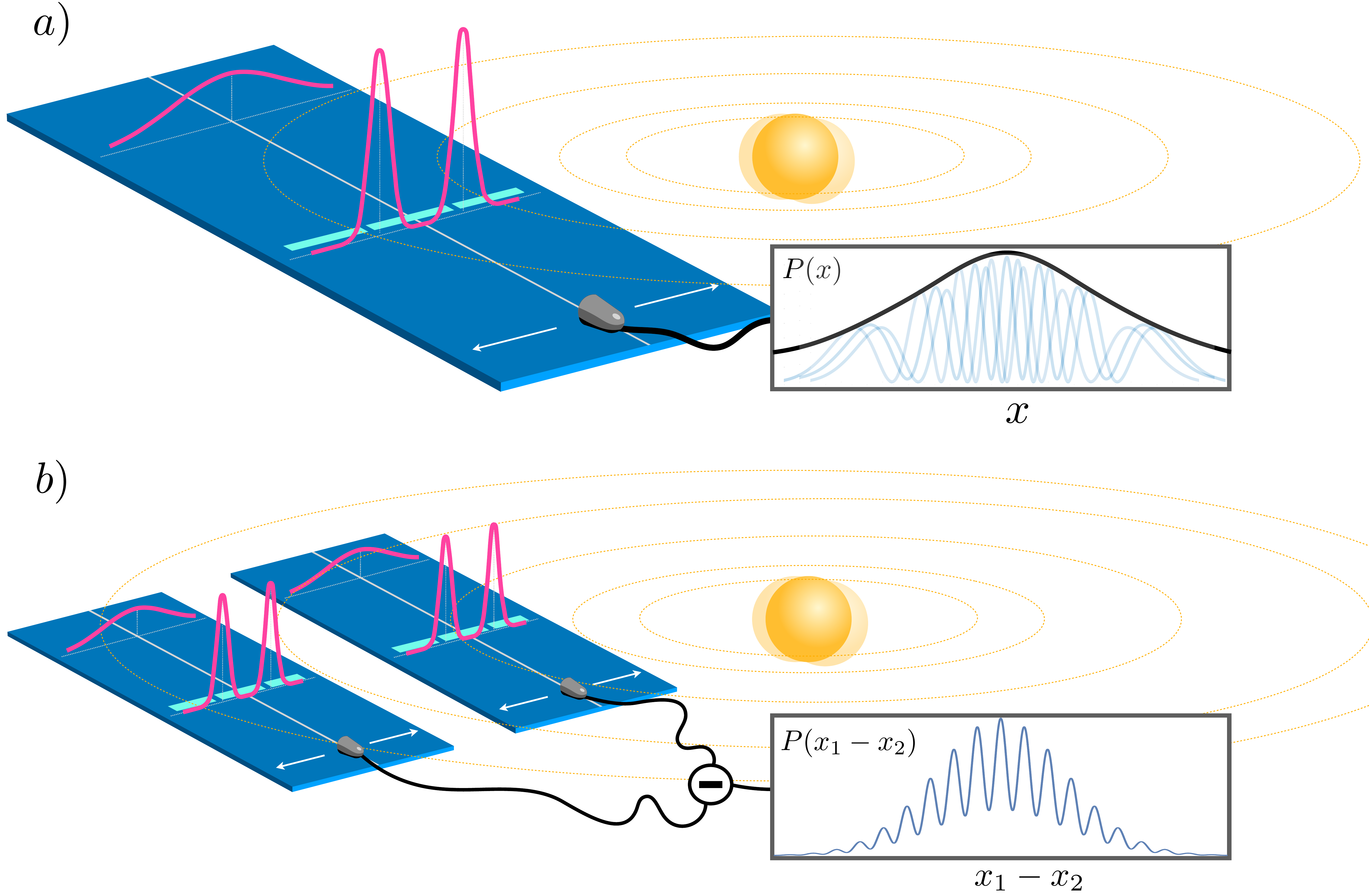} 
   \caption{{\bf Single and doubled matter-wave interferometry.} a) shows a generic matter-wave interferometer, where a massive object is prepared in a pure superposition state, consisting of two Gaussian wave packets separated in space. After some time, the position of the systems is measured. The probability distribution of the particle position exhibits an interference pattern. However, in the presence of a fluctuating mass density (yellow sphere) of uncontrolled origin, the interference pattern will fluctuate from shot-to-shot repetition of the experiment and may be lost upon averaging. b) shows our proposed setup consisting of two interferometers subject to the same source of noise. The individual probability distributions for each of the interferometers are shifted in a correlated manner, such that the probability distribution of the difference between the position of each interferometer is independent of such fluctuations and contains a robust interference pattern in its probability distribution. }
   \label{fig:setup}
\end{figure}

As opposed to matter-wave interferometers based on ensembles or beams of a large number of identical particles, each run of an optomechanical device will typically operate with a single particle. This implies that in each run of the experiment a unique single body, whose properties may change from shot-to-shot, needs to be initialized and probed to reconstruct the probability distribution in which the interference pattern of interest is encoded: typically, but not necessarily, that corresponding to the position of the c.m. The faithful reconstruction of a probability distribution that encodes an interference pattern from sequentially repeated experiments requires that every experimental run prepares the system in the same quantum state prior to measurement, thus ensuring that one is not averaging over different probability distributions. Noise in the form of an interferometric phase that fluctuates from shot-to-shot of the experiment, will shift the interference pattern between measurements, and destroy any observable interference if the distribution fluctuates by an amount that exceeds the distance between fringes, see Fig~(\ref{fig:setup}).   

Therefore, not only will a solid-state matter-wave interferometer be more sensitive to noise on its phase due to its increased mass, but it requires that the stability of such a phase is preserved over longer periods of time, in order to allow for the observation of interference in the reconstructed probability distribution. This represents an important limitation for these type of devices, in particular due to the presence of acceleration noise, including that of gravitational origin, from which experiments cannot be shielded easily \cite{PMP20}. 

In this Letter, we introduce a framework by which a generic matter-wave interferometer can be made robust to field fluctuations of uncontrolled origin up to an arbitrary order. We show that an interference pattern can be recovered from the joint probability distribution of two or more interferometers, even when the noise is such that the interference pattern is erased from each of them individually. Our formalism accounts for the possibility that the interferometers have differing masses, and is thus well suited for interferometry with solid state objects. Moreover, we do not require that the interferometers are entangled. 

{\it Single-device interferometry---}A matter-wave interferometer is a device that can generate a coherent quantum superposition of some mechanical degree of freedom of a massive object and detect an interference pattern in its probability distribution. Without loss of generality, we consider an interferometer that prepares the c.m.\@ of a solid of mass $m$ in a symmetric superposition of two coherent states $\ket{\Psi}_{\rm in} = (\ket{\alpha} + \ket{-\alpha})/N_\alpha$, with $N_\alpha$ a normalization constant, and $\alpha$ a complex number that we assume can be chosen at will (see~\cite{PPS+18}
for a possible realisation). Here, the coherent states are given with respect to some reference mode of frequency $\omega$, which allows us to parametrize the width, position and momentum of the superposed wave packets through the values of $\omega$ and $\alpha$ alone. The particular way in which such an initial state is prepared is not the concern of this paper, and the results presented here are independent of it. We are interested in the evolution of such a state when continuously subjected to an acceleration $g(t)$, which in the most general case is a stochastic function of time that might fluctuate slower or faster than the duration of a single experimental run. One natural and difficult to shield origin of such an acceleration is the presence of a fluctuating gradient of the gravitational potential, which is to be expected in all matter-wave interferometers that are not performed in free fall. We are concerned with the detrimental effects of such an acceleration noise on the visibility of the interference pattern after some evolution time $t$, when averaged over many experimental runs and in ways to mitigate it. To that end, we compute the evolution of the system under a Hamiltonian of the form $\hat H = \frac{\hat p^2}{2 m} + m g(t) \hat x$, where the second term accounts for the acceleration noise, and then consider the probability distribution of the position of the c.m. $\hat x = x_0 (\hat a + \hat a^\dag)$, with $x_0= \sqrt{\hbar/(2m\omega)}$. With the normalization constant $N_t$ the result is~\cite{SuppMat}
\begin{multline}
\label{eq:ProbDis}
P(x,t) = \frac{1}{N_t} ( e^{- \frac{[x - x_\gamma(t) - x_{\alpha}(t) ]^2}{2\sigma_t^2} } + e^{- \frac{[x - x_\gamma(t) - x_{-\alpha}(t) ]^2}{2 \sigma_t^2}} \\
+ 2 e^{- \frac{x_{\alpha}(t)^2}{2\sigma_t^2}} e^{- \frac{[x - x_\gamma(t)]^2}{2 \sigma_t^2}} \cos{ \{k_t [x - x_\gamma(t)]\}}  ).
\end{multline}
Here, $x_{\pm \alpha}(t) = \pm 2 x_0 ( \alpha_r + \alpha_i \omega t)$ give the linear motion of the centers of the wave packets under free evolution according to their initial position and momentum, determined by the real, $\alpha_r$ and imaginary, $\alpha_i$, components of $\alpha$. At the same time, the width of the wave packets spreads linearly in time with $\sigma_t^2 = x_0^2 [1 + (\omega t)^2]$. Finally, the interference term exhibits an oscillatory pattern characterized by a wave number $k_t = 2/x_0 \{\alpha_i + \omega t / [1 + (\omega t)^2] (\alpha_r + \alpha_i \omega t)\}$ and an envelope determined by the overlap between the two Gaussian wave packets.   The effect of the acceleration noise is to stochastically displace the probability distribution in the position variable by an amount $x_\gamma (t) = \int_0^t g(s) (s-t) ds$.
The induced displacement is independent of mass or initial state of the interferometer and depends only on time and the particular realization of the stochastic function $g(t)$ during the evolution. For $\alpha_i\neq 0$, we find that at time $t_k = - \alpha_r/(\omega \alpha_i)$, $x_{\pm \alpha} = 0$ and the two wave packets overlap completely in space, resulting in a probability distribution with the simpler form
\begin{equation}
\label{eq:ProbDisSimp}
P(x, t_k) = \frac{4}{N_{t_k}} e^{- \frac{ [x - x_\gamma(t_k) ]^2}{2 \sigma_{t_k}} } \cos^2{\{ \frac{k_{t_k}}{2} [x - x_\gamma(t_k)]\}},
\end{equation}
with $k_{t_k} = 2 \alpha_i/x_0$. In the interest of simplicity and clarity of the presentation, for the remainder of this work, we shall work with the probability distribution at this point in time, but all the results can be derived with the same methods using the distributions corresponding to any other time. 

In order to reconstruct the probability distribution in Eq.~(\ref{eq:ProbDisSimp}) experimentally, a statistically significant number of experimental runs will be required.  Shot-to-shot fluctuations of $x_\gamma(t_k)$ will amount to sampling from a different probability distribution in each experimental run, and thus, in a loss of visibility of the interference pattern. In particular, when the fluctuations extend over distances greater than the separation between fringes, $2/k_t$, the interference pattern will be lost. If we assume that in each experimental run $x_\gamma (t_k)$ takes a random value with a normal probability distribution $\mu(x_\gamma) = 1/(\sqrt{2 \pi} \sigma_\gamma) \exp{-x_\gamma^2 /(2 \sigma_\gamma^2)}$, characterized by a standard deviation $\sigma_\gamma$, the averaged probability distribution $\bar P(x) = \int_{-\infty}^\infty dx_\gamma \mu(x_\gamma) P(x,t_k)$ takes the form
\begin{equation}
\bar P(x) = \frac{e^{- \frac{x^2}{2(\sigma_{t_k}^2 + \sigma_\gamma^2)}}}{\bar N}  [ 1 + e^{- k^2 \frac{\sigma_{t_k}^2 \sigma_\gamma^2}{2(\sigma_{t_k}^2 + \sigma_\gamma^2)}} \cos{ (k_{t_k} \frac{\sigma_{t_k}^2}{\sigma_\gamma^2 + \sigma_{t_k}^2} x) } ],
\end{equation}
with $\bar N$ a normalization constant. In the limit where the fluctuations are smaller than the width of the wave packet at the time of measuring, that is when $\sigma_\gamma \ll \sigma_{t_k}$, the oscillatory term in the probability distribution is attenuated by a factor $\exp{- k^2 \sigma_\gamma^2/2}$, which shows that the interference pattern is quickly suppressed as soon as the fluctuations exceed the distance between fringes. Similarly, in the limit $\sigma_{\gamma}\gg \sigma_{t_k}$, the interference term is suppressed by a factor $\exp{- k^2 \sigma_{t_k}^2/2}$, as we are in the regime where the envelope of the Gaussian wave packet exceeds the periodicity of the fringes $1/k$.

{\it Two-device interferometry.---}We now consider two interferometers, 1 and 2, of the type described in the previous section, which are operated simultaneously and in parallel at a distance close enough from each other that they can be considered to be affected by the same acceleration noise $g(t)$. At time $t_k$, the outcome of a position measurement is described in each interferometer by a probability distribution of the form of Eq.~(\ref{eq:ProbDisSimp}), $P_1$ and $P_2$, with parameters $\sigma_t$ and $k_t$ determined by the mass of each interferometer, $m_1$ and $m_2$, which in the general case will differ from each other. While the form of each probability distribution will be different, the fluctuations of the local position variables in each of the two distributions will be the same. This is because, as mentioned earlier, $x_\gamma$ depends only on the acceleration function $g(t)$ and time, which here we assume to be the same for both interferometers. Thus, the joint probability distribution $P_{12}(x_1,x_2)= P_1(x_1)P_2(x_2)$ is a two-variable function that fluctuates in the direction of $x_1 + x_2$. Therefore, the dependence of the joint probability distribution on the stochastic variable $x_\gamma$ can be removed by the variable transformation $x_\pm \rightarrow (x_1 \pm x_2)/2$ and integration over $x_+$. 

\begin{figure}[htbp]
   \centering
   \includegraphics[width=\columnwidth]{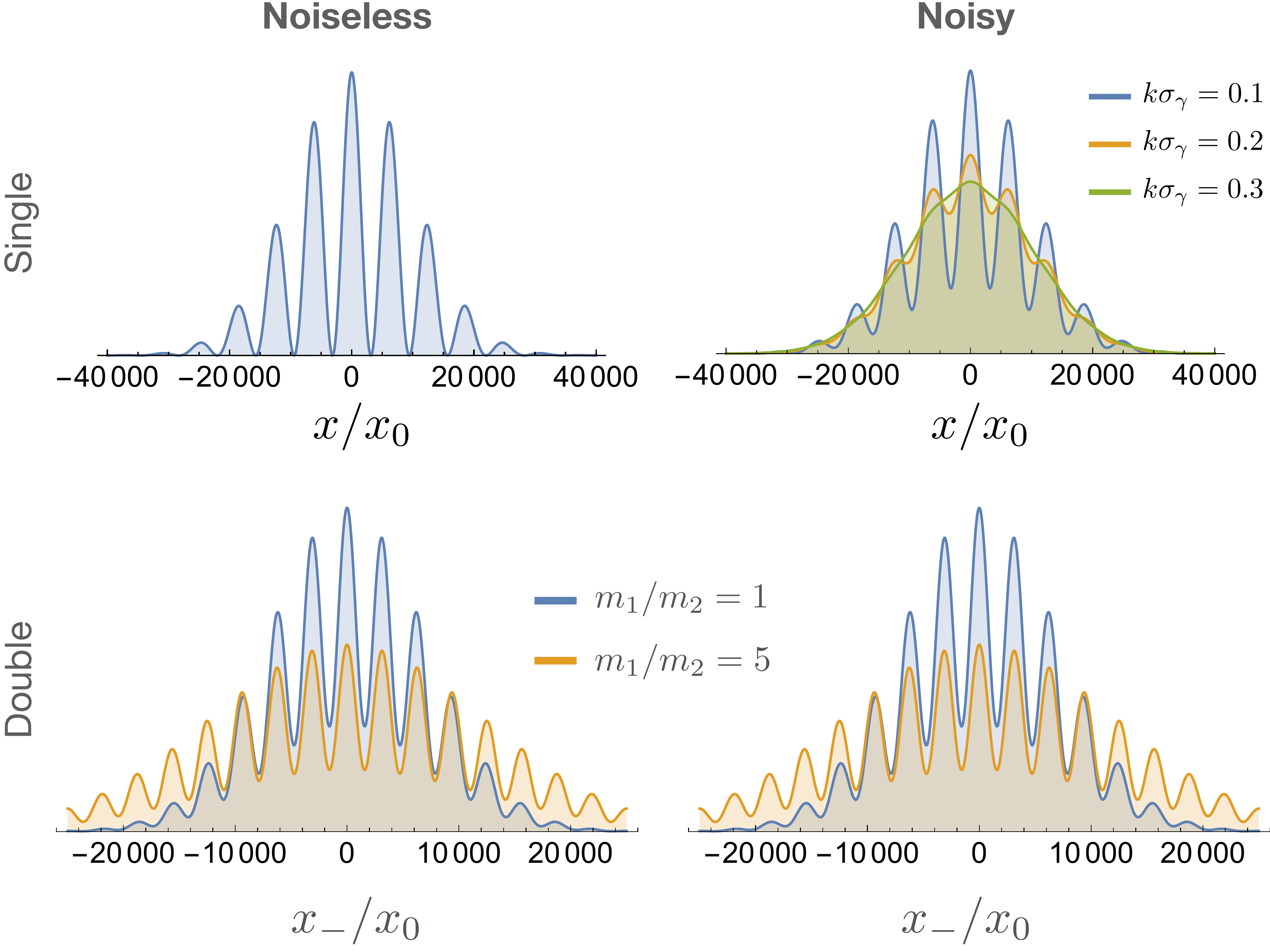} 
   \caption{{\bf Interference pattern.} The top left panel shows the probability distribution of the position of a single-particle interferometer in the absence of noise sources, as given by Eq.~(\ref{eq:ProbDis}) with $x_\gamma (t) = 0$. Here, the parameters where chosen as $k x_0 = 10^{-3}$ and $\sigma_{t_k} = 10^4 x_0$ with illustrative purpose. On the top right, we show the same interference pattern but now in the presence of an acceleration noise characterized by a variance $\sigma_\gamma$ for the stochastic variable $x_\gamma$.  The lower row shows the same two scenarios now for the probability distribution of $x_-$ in a system of two interferometers. We compare the cases of equal masses and masses with a ratio $m_1/m_2=5$.}
   \label{fig:interference}
\end{figure}

However, while the resulting probability distribution for the variable $x_-$, $P_- (x_-)= \int_{-\infty}^\infty dx_+P_{12}(x_+,x_-)$, has been freed of any stochasticity, it might not retain an interference pattern. From the trigonometric identity $\cos{A}\cos{B} = \cos(A + B) + \cos{(A-B)}$, we see that the joint probability distribution $P_{12}$ contains an oscillatory term of the form $\cos{ [k_+ x_- + k_- (x_+ - x_\gamma)]}$, with $k_\pm = (k_1 \pm k_2)/2$. For the case in which $k_1= k_2 = k$ this term is independent of $x_+$ and thus at least one interferometric term will survive after integration over $x_+$.  Given that, for each interferometer, $k \propto \abs{\alpha} \sqrt{m}$, such a condition can be ensured provided that the interferometers are initialized in cat states with a choice of $\alpha$ that keeps the product $\abs{\alpha}\sqrt{m}$ equal for both interferometers. Notice that this uneven initialization does not affect the time $t_k$ at which the probability distributions adopt the form of Eq.~(\ref{eq:ProbDisSimp}), as this time depends only on the ratio $\alpha_r/\alpha_i$ and $\omega$, which can be maintained the same in both interferometers. Thus, for suitably initialized interferometers, we find that the probability distribution of variable $x_-$ is given by
\begin{equation}
\begin{split}
\label{eq:PatDoub}
P_-(x_-) = \frac{ e^{- \frac{x_-^2}{2\sigma_+^2}}}{N_-}\{ 2 + \cos{(2 k x_-)}
+ 2 \eta [ \cos{(\frac{k \sigma_a^2}{2 \sigma_+^2} x_-)} \\ +  \cos{(\frac{k\sigma_b^2}{2 \sigma_+^2} x_-)} ] + \eta^4 \cos{ (\frac{k \sigma_-^2 }{2 \sigma_+^2} x_- )} \},
\end{split}
\end{equation}
where $N_-$ is a normalization constant, and we have defined $\sigma_\pm^2 = (\sigma_1^2 \pm \sigma_2^2)/4$ and $\eta = \exp{-k^2 \sigma_1^2 \sigma_2^2 / (8 \sigma_+^2)}$. In general, we will have that $k \sigma_{1(2)} \gg 1$ which ensures that both interferometers exhibit an interference pattern. It follows, that for all cases of interest $\eta \ll 1$ and we can safely neglect in Eq.~(\ref{eq:PatDoub}) terms proportional to $\eta$ or higher orders of it. The interference pattern of the variable $x_-$ is thus dominated by a term of the form $\exp{-x_-^2/(2\sigma_+^2)} [2 + \cos(2 k x_-) ]$. The price to pay for the gained robustness against fluctuations of the acceleration is a reduction in visibility of the interference pattern. We observe that the width of the Gaussian envelope is reduced by a factor $1/\sqrt{2}$ and that the frequency of the pattern is doubled. At the same time, the second term in Eq.~(\ref{eq:PatDoub}) reduces the distance between the maxima and minima of the interference fringes.  

In Fig.~(\ref{fig:interference}), we confirm that in the presence of noise that is sufficient to erase the interference pattern in individual interferometers, the doubled interferometer system in a differential configuration, $x_-$, exhibits an interference pattern that is insensitive to phase fluctuations. Remarkably, this remains true for any time dependency of the acceleration field, including fluctuations that occur in a time scale much smaller than the duration of each experimental run.

\begin{figure}[htbp]
   \centering
   \includegraphics[width=\columnwidth]{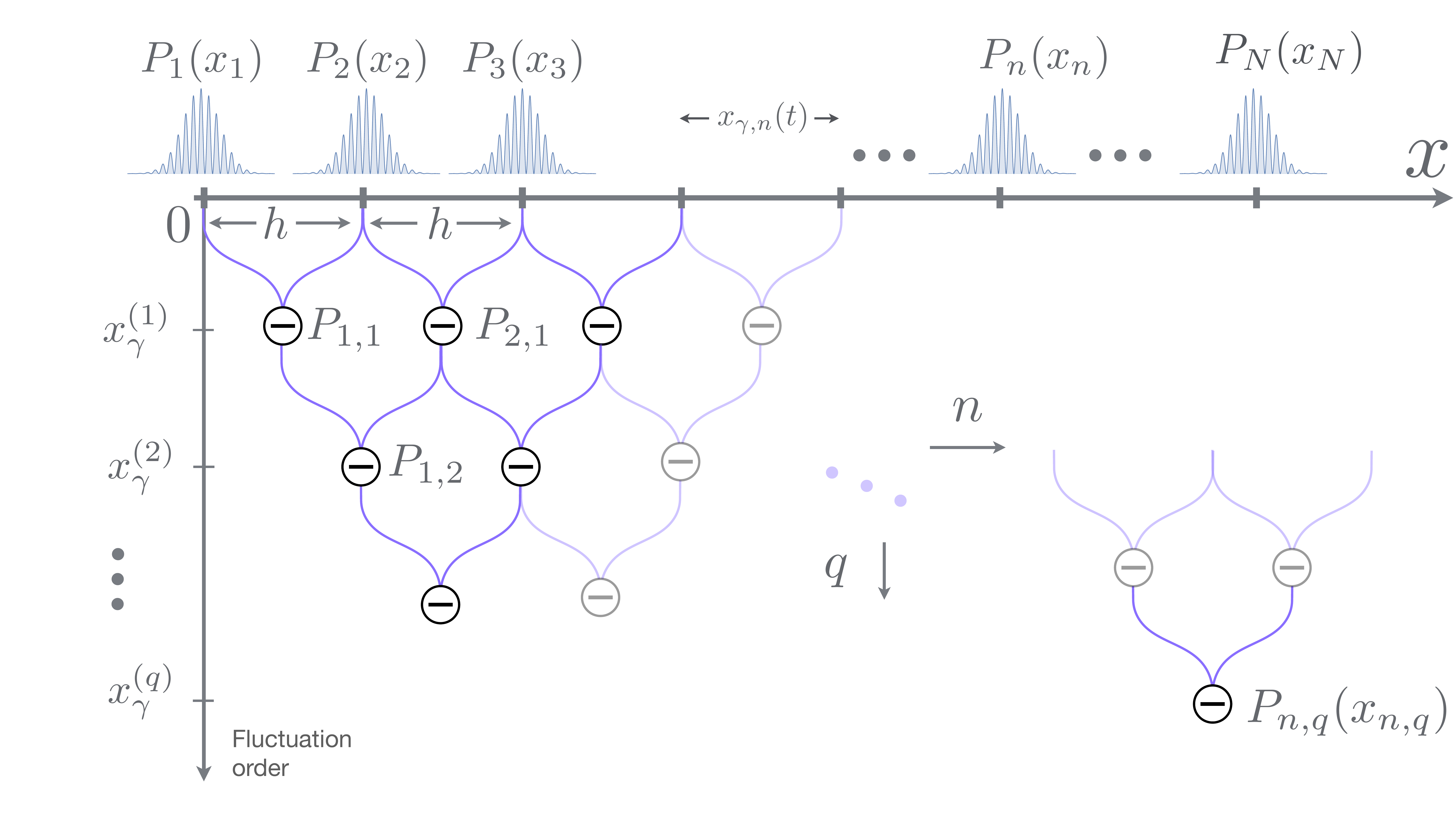}
   \caption{{\bf Higher-order interferometry.} Array of $N$ interferometers disposed along the x-axis and spaced by a distance $h$, where interferometer $n$ displays a probability distribution $P_n(x_n)$ subject to a stochastic displacement of its local position variable $x_n$ by an amount $x_{\gamma, n}(t)$. Variables $x_{n,q}$ with probability distribution $P_{n,q}$ that are subject only to noise of order $x_{\gamma,n}^{(q)}$ can be constructed by iteratively subtracting two consecutive variables of one lower order.}
   \label{fig:higherOrder}
\end{figure}

{\it Higher orders in the multipole expansion.---}So far we have considered the case of two interferometers subject to the same acceleration noise. We now extend this concept to an array of $N$ interferometers subject to acceleration noise that might differ at the location of each interferometer, and show that an interference pattern can be reconstructed that is robust against inhomogeneous acceleration noise up to order $N$ of its expansion in the ratio of the spatial extent of the array to the distance from the source of the perturbing field. We consider that the $N$ interferometers are disposed in a linear arrangement along dimension $x$, and that they are evenly spaced by a distance $h$, see Fig.~ (\ref{fig:higherOrder}). We expand the acceleration field around the origin of coordinates, which we conveniently place at the position of the first interferometer,
\begin{equation}
g (x, t) = g^{(0)}(t) + g^{(1)}(t) x + g^{(2)} (t) x^2 + \dots
\end{equation}
The local position variable $x_n$ at interferometer $n$ fluctuates with $x_{\gamma,n} (t) = \int_0^t g(n h, s) (s-t) ds $, which we can express in orders of the expansion of the acceleration field as
\begin{equation}
x_{\gamma,n} (t) = x_\gamma^{(0)}(t) + n h x_\gamma^{(1)} (t) + (n h)^2 x_\gamma^{(2)} \dots,
\end{equation}
where the coefficients $x_{\gamma}^{(k)} (t) =  \int_0^t g^{(k)}(s) (s-t) ds$ fluctuate in time and are independent of the position of the interferometer. In the previous section, we have shown that a variable constructed from the difference between the positions of two consecutive interferometers $x_{n,1} = (x_{n} - x_{n+1})/2$ is independent of zero-order fluctuations $x_\gamma^{(0)}$, and fluctuates with the next order as $h x_\gamma^{(1)}$. Since $x_{n,1}$ is subject to fluctuations which to the first significant order are independent of $n$, this fluctuations can be removed by subtracting two such variables. That is, three consecutive interferometers can be used to construct a second-order variable $x_{n,2} = (x_{n,1} - x_{n+1,1})/2 = (x_{n} - 2 x_{n+1} + x_{n+2})/4$ which is sensitive only to fluctuation of order $h^2 x_\gamma^{(2)}$. In terms of $x_{\gamma, n}(t)$, this linear combination of positions is the discretized second derivative with respect to $n$. Hence, terms of order zero and one are cancelled while quadratic and higher-order terms survive. More generally, $q+1$ interferometers can be used to construct a variable $x_{n,q}$ of order $q$ that corresponds to the $q$-th order derivative w.r.t. $n$ and is therefore only sensitive to fluctuations of order $q$, which have the form $q! h^q x_\gamma^{(q)}/2^q$. Such a variable can be constructed recursively by subtracting two consecutive variables of order $q-1$, that is $x_{n,q} = (x_{n,q-1} - x_{n+1,q-1})/2$, and it has the general expression
\begin{equation}
    x_{n,q} = \sum_{i=0}^q (-1)^i \frac{q!}{2^q i! (q-i)!} x_{i+n}.
\end{equation}
We are therefore interested in checking whether the probability distribution of variables $x_{n,q}$ exhibit an interference pattern. 

In general, the probability distribution of the difference of two stochastic variables $x_- = (x_1 - x_2)/2$, each described by a probability distribution of the form
\begin{equation}
\label{eq:GenerProbDist}
P(x) = \frac{1}{N} e^{-\frac{x^2}{2 \sigma}} [a + \cos(k x)],
\end{equation}
is obtained by the convolution $P_- (x_-) = \int_{-\infty}^\infty dx_2 P_1(2 x_- + x2) P_2 (x_2)$ and has the explicit expression
\begin{multline}
 P_- (x_-) =  \frac{1}{N_-} e^{-\frac{x_-^2}{2\sigma_+^2}}  \{ 2 a^2 + \cos{(2 k x_-)} + 2 \eta a [ \cos{(\frac{\sigma_1^2}{2\sigma_+^2} k x_-)} \\ + \cos{(\frac{\sigma_2^2}{2\sigma_+^2} k x_-)} ] + \eta^4 \cos{(\frac{\sigma_1^2 - \sigma_2^2}{2\sigma_+^2} k x_-)} \},
\end{multline}
with $\eta =  \exp{-k^2 \sigma_n^2 \sigma_{n+1}^2 / (8 \sigma_+^2)}$, $\sigma_+^2 = (\sigma_{n+1}^2 + \sigma_n^2)/4$. As discussed in the previous section, we can discard terms proportional to $\eta$ on the basis that $k \sigma_{1(2)} \gg 1$, which leaves us with a probability distribution of the same functional form as that of Eq.~(\ref{eq:GenerProbDist}). We can use this fact to construct recursively probability distributions for variables $x_{n,q}$ of order $q$,  from two probability distributions of a previous order. Thus, we have that the probability distribution for $x_{n,q}$ is given by
\begin{equation}
P_{n,q} (x_{n,q}) = \frac{1}{N_{n,q}} e^{-\frac{x_{n,q}}{2 \sigma_{n,q}^2}} [ a_q + \cos(k_q x_{n,q}) ],
\end{equation}
where the parameters can be computed from the parameters of the previous order using the following update formulas
\begin{equation}
 a_{q+1} = 2 a_q^2 \qquad \sigma_{n,q+1}^2 = \frac{ \sigma_{n,q}^2 + \sigma_{n+1,q}^2}{4} \qquad  k_{q+1} = 2 k_q.
\end{equation}

We observe that a trade-off exists between robustness and visibility of the interference patterns as we move into higher orders. While with every order we become insensitive to one higher order of the acceleration field, the visibility of the interference pattern is decreased in several ways. On the one hand, the frequency of the oscillations is doubled with each order, which makes the fringes more difficult to distinguish. On the other hand, the constant $a_q$, which determines the amplitude of the oscillations, is squared with every iteration. Moreover, the width of the Gaussian envelope determined by $\sigma_{n,q}$ is reduced approximately by a factor of $1/\sqrt{2}$ with every order.  

In order to illustrate the power of using multiple interferometric devices, we consider an experimental situation where acceleration noise below $\Delta a = 6.67 \cdot 10^{-17}$ m/s$^2$ is required~\cite{PPS+18}. In such a scenario, a single interferometer would be significantly perturbed by a mass variation of $1$ Kg at a distance of $1$ Km, an effect that could be caused, for example, by 1 ppb pressure or temperature fluctuations of an atmospheric environment. Our proposed system of two interferometers, however, would only be sensitive to the relative acceleration, which if we chose to place the two interferometers at a distance of $h=0.1$ m, would require the presence of the same $1$ Kg mass at the much closer distance of $R = 58$ m to produce a comparable perturbation in the interference pattern. Extending our approach to three setups we would find that the distance is further reduced to $R\cong 15$ m, which is a volume that is reasonable to expect could be brought under experimental control. 

It is noteworthy to mention that the principles employed here can be extended to experiments where the interest is not in protecting the coherence of a single interferometer from acceleration noise, but rather the entanglement between two such interferometers, see~\cite{SuppMat}.

{\it Conclusion.---}We introduce a general theory to construct arrays of $N+2$ matter-wave interferometers that can combat the effects of acceleration noise up to order $N$ of the expansion of the acceleration field along the dimension of the array. In this manner, we can recover interference patterns that are otherwise erased by this form of noise. While the visibility of higher-order interference patterns is reduced, the available experimental time to reconstruct the probability distribution becomes unlimited as our strategy cancels noise fluctuating at any time scale. Our design is general and considers that the different interferometers may differ in terms of their mass and that they are placed in different positions of space. Moreover, we assume that the interferometers are operated independently, and thus do not require to be entangled, as opposed to other noise mitigation techniques with duplicated systems that work by constructing decoherence-free subspaces.  These aspects, make our framework particularly suited to matter-wave interferometry using solid-state objects, where the replication of redundant interferometers necessarily involves differing masses and locations, as opposed to interferometers that use ensembles of identical particles. All in all, this introduces a framework for the systematic construction of robust macroscopic matter-wave interferometers and brings the possibility of interferometry with solid-state objects closer to reality. 

\section*{Acknowledgments} This work was supported by the ERC Synergy grant HyperQ (Grants No. 856432), the EU projects and AsteriQs (Grants No. 820394) and the BMBF project Q.Link.X (Contract No. 16KIS0875).

\begin{widetext}
\newpage

\section{Supplemental Material:\\ Robust Macroscopic Matter-Wave Interferometry with Solids}

\subsection{Derivation of the probability distribution}
\label{appendix:ProbDis}

In this section we compute the probability distribution of the position of an initial ``cat'' state $\ket{\psi}_{\rm in} = 1/N_\alpha (\ket{\alpha} + \ket{-\alpha})$ that evolves freely in the presence of a fluctuating gradient of the gravitational field, governed by the Hamiltonian
\begin{equation}
\label{equationApp:Ham}
H_A = \frac{\hat p^2}{2m} + m g(t)\hat x.
\end{equation}

We first set the notation, $\hat x = x_0 (a + a^\dag)$ and $\hat p = ip_0(a^\dag - a)$, where $a$ is defined such that the components $\ket{\alpha}$ of the initial state are its eigenstates, $a \ket{\alpha} = \alpha \ket{\alpha}$. Notice that $x_0$ and $p_0$ correspond, respectively, to the position and momentum variances of the coherent state ($A_0^2 = \bra{\alpha}\hat A^2\ket{\alpha} - \bra{\alpha}\hat A \ket{\alpha}^2$, with $A =  x,  p$) and satisfy the relation $x_0p_0=\hbar/2$. Under these definitions the free-evolution Hamiltonian~(\ref{equationApp:Ham}) can be rewritten as
\begin{equation}
H_A = \frac{p_0^2}{2m} [ 2a^\dag a - ({a^\dag}^2 + a^2) ] + m g(t) x_0 (a + a^\dag).
\end{equation}
By defining $H_0 = \frac{p_0^2}{2m} [ 2a^\dag a - ({a^\dag}^2 + a^2) ] $, the time evolution operator can be expressed as $U(t- t_0) = U_0(t-t_0)U_I(t-t_0)$, where $U_I(s)$ is the time-evolution operator in the interaction picture with respect to $H_0$ and $U_0(s) = \exp{(- i/\hbar) H_0 s}$. To give an explicit expression for $U_I(s)$, we first derive the Hamiltonian in the interaction picture
\begin{equation}
H_I (s) = m g(s) x_0 [(1-i\frac{2p_0^2}{m\hbar}s) a + (1 + i \frac{2p_0^2}{m\hbar}s) a^\dag],
\end{equation}
where we have used the transformation $U_0^\dag a U_0 = a + \frac{it}{\hbar} [H_0,a] + \frac{1}{2!} \frac{(it)^2}{\hbar^2} [H_0,[H_0,a]] + \dots$, $[H_0, a] = \frac{p_0^2}{m} (a^\dag - a)$ and the fact that all higher-order nested commutators vanish identically. The unitary-evolution operator associated to the time-dependent Hamiltonian $H_I(s)$, which does not commute at different times, can be exactly computed in the Magnus expansion
\begin{equation}
U_I (t) = e^{\Omega^{(1)}(t) + \Omega^{(2)}(t) + \dots},
\end{equation}
with
\begin{align}
\Omega^{(1)}(t) &= \gamma_t a^\dag - \gamma_t^* a \\
\Omega^{(2)}(t) &= - \frac{i 2 p_0^2}{\hbar^3} \int_0^t dt_1 \int_0^{t_1} dt_2 g(t_1)' g(t_2)' (t_2 - t_1),
\end{align}
where $\gamma_t = - \frac{i m x_0}{\hbar} \int_0^t ds g(s) (1 + i\frac{2p_0^2}{m\hbar}s)$, and all higher-order terms vanish. The first term represents a displacement by the complex time-dependent amplitude $\gamma_t$ in phase space, while the second order is just a global phase that can be ignored. Back in the Schr\"odinger picture the time evolution is given by
\begin{equation}
U(t) = e^{-\frac{{i\hat p}^2}{2 m \hbar} t} e^{\gamma_t a^\dag - \gamma_t^* a}.
\end{equation}
We look, now, into the action of this operator on the initial state
\begin{equation}
U(t) \ket{\psi}_{\rm in} = \frac{1}{N_\alpha} U_0(e^{i \phi(t)}\ket{\alpha + \gamma_t} + e^{-i\phi(t)}\ket{-\alpha + \gamma_t}),
\end{equation}
where 
\begin{align}
\phi(t) &= \Im{\gamma_t \alpha^*} \\
&= - \frac{m x_0 \alpha_{\rm r}}{\hbar} \int_0^t ds g(s) + \frac{2 p_0^2 x_0 \alpha_{\rm i}}{\hbar^2} \int_0^t ds g(s) s
\end{align} 
with $\alpha = \alpha_{\rm r} + i \alpha_{\rm i}$. Here, we have used the composition rule for the displacement operator $D(\beta) D(\alpha) = e^{(\beta \alpha^* - \beta^*\alpha)/2} D(\alpha + \beta)$. Ultimately, we are interested in the probability distribution $P(x) = \abs{\langle x | \psi (t) \rangle}^2$ which takes the form
\begin{align}
\label{equationAppendix:ProbDis}
P(x) &= \frac{1}{N_\alpha^2} ( \abs{\bra{x} U_0\ket{\alpha + \gamma_t}}^2 +\abs{\bra{x} U_0\ket{-\alpha + \gamma_t)}}^2 \\
&+ (\bra{x} U_0\ket{\alpha + \gamma_t}\bra{-\alpha + \gamma_t} U_0^\dag \ket{x} e^{i2\phi(t)} + {\rm H.c.}).
\end{align}
The terms $\bra{x} U_0 \ket{\alpha}$ are time dependent and correspond to the wave functions in the spatial representation of a coherent state evolving under free-evolution. Their explicit expression is
\begin{equation}
\label{equationApp:cohFree}
\bra{x} U_0 \ket{\alpha} = \frac{1}{\sqrt{1 + i \frac{2p_0^2}{\hbar m}t }} \left( \frac{1}{2 \pi x_0^2} \right)^{1/4} \exp{-\frac{[x - (\expval{x}_\alpha + \frac{\expval{p}_\alpha}{m}t) ]^2}{4 x_0^2 [1 + (\frac{2p_0^2}{\hbar m}t)^2] } (1 - i\frac{2 p_0^2}{\hbar m} t)} \exp{i\frac{\expval{p}_\alpha}{\hbar}[x - (\expval{x}_\alpha + \frac{\expval{p}_\alpha}{m}t)/2]},
\end{equation}
where $\expval{x}_\alpha = 2 x_0 \alpha_{\rm r}$ and $\expval{p} = 2 p_0 \alpha_{\rm i}$ are, respectively, the position and momentum expectation values of the initial state $\ket{\alpha}$. Finally, inserting Eq.~(\ref{equationApp:cohFree}) in Eq.~(\ref{equationAppendix:ProbDis}) we recover the probability distribution
\begin{equation}
P(x,t) = \frac{1}{N_t} \left( e^{- \frac{[x - x_\gamma(t) - x_{\alpha}(t) ]^2}{2\sigma_t^2} } + e^{- \frac{[x - x_\gamma(t) - x_{-\alpha}(t) ]^2}{2 \sigma_t^2}} \\
+ 2 e^{- \frac{x_{\alpha}(t)^2}{2\sigma_t^2}} e^{- \frac{[x - x_\gamma(t)]^2}{2 \sigma_t^2}} \cos{ \{k_t [x - x_\gamma(t)]\}}  \right).
\end{equation}

\subsection{Entanglement}
\label{appendix:Spins}

In this section, we discuss the case where the primary entity of interest is not one interferometer but a system composed of two entangled interferometers, and we analyse one possible application of the same principles discussed in the main text for the protection of a single-device coherence to the protection of the entanglement between these two devices.

In this case, a single device consists of two interferometers $A$ and $B$, where each interferometer can prepare states $\ket{L}$ (for the left arm) and $\ket{R}$ (for right arm) in superposition. We are interested in the entanglement between $A$ and $B$ in this two-dimensional subspace of $\{ \ket{L}, \ket{R} \}$. For simplicity, we will consider maximally entangled states of the form $ 1/\sqrt{2} (\ket{L}_A\ket{L}_B + e^{i \phi}\ket{R}_A \ket{R}_B)$, where the phase $\phi$ represents our source of noise, which appears due to the presence of a gradient of a potential field (e.g. gravitational) and will, in general, fluctuate randomly in time scales shorter than the duration of an experimental run. If $\phi$ has a variance $\sigma_\phi > 2 \pi$, then the entanglement in our double-interferometer setup will be lost upon averaging over different realization of $\phi$, such that the state of the system will be given by
\begin{equation}
    \expval{\rho}_\phi = \frac{1}{2} (\mathds{P}_{\ket{L}_A\ket{L}_B} + \mathds{P}_{\ket{R}_A \ket{R}_B}),
\end{equation}
where $\mathds{P}_{\ket{\psi}} = \ket{\psi}\bra{\psi}$ is the projector on the state indicated in its subscript. 

We now double our device and check if entanglement survives in some bi-partition of the 4 interferometers upon averaging over the stochastic source of noise. In this case, the total state of the system before averaging is given by a tensor product of the two entangled states of each of the interferometers $1$ and $2$, that is
\begin{equation}
\label{eq:pairconst}
    \Psi = \frac{1}{2}\left( \ket{L}_{A1}\ket{L}_{B1} + e^{i \phi}\ket{R}_{A1} \ket{R}_{B1}\right) \otimes \left( \ket{L}_{A2}\ket{L}_{B2} + e^{i \phi}\ket{R}_{A2} \ket{R}_{B2} \right),
\end{equation}
where, for the moment, we are assuming that both devices observe the same form of noise, and therefore, pick up the same random phase $\phi$. Averaging over this phase yields the mixed state
\begin{equation}
\label{eq:doubleavg}
    \expval{\rho}_\phi = \frac{1}{4} \mathds{P}_{\ket{LL}_A \ket{LL}_B} + \frac{1}{4} \mathds{P}_{\ket{RR}_A \ket{RR}_B} +\frac{1}{2} \mathds{P}_{(\ket{LR}_A \ket{LR}_B + \ket{RL}_A \ket{RL}_B)/\sqrt{2}},
\end{equation}
where we have introduced the abbreviated notation $\ket{ij}_\alpha \equiv \ket{i}_{\alpha 1} \ket{j}_{\alpha 2}$, with ${i,j \in \{ L, R \}}$ and ${\alpha \in \{A,B\}}$. Remarkably, the state in Eq.~(\ref{eq:doubleavg}) contains half a bit of entanglement between the system composed of interferometers $A$ in devices $1$ and $2$, and the system composed of interferometers $B$ as can be confirmed by calculation of the logarithmic entanglement of the state \cite{Plenio2005}. Apart from direct computation, another way to see state~(\ref{eq:doubleavg}) is entangled is the following. Consider a local observable in the partition $A$ of the total system which is of the form
\begin{equation}
    \mathcal{O}_A = \ketbra{LL}{LL} + 2 (\ketbra{LR}{LR} + \ketbra{RL}{RL}) + 3 \ketbra{RR}{RR}
\end{equation}
Measurement of this observable on state~(\ref{eq:doubleavg}) will yield outcome $2$ with probability $p=1/2$ and leaves the system in the maximally entangled state
\begin{equation}
\label{eq:entangled}
    \rho = \mathds{P}_{(\ket{LR}_A\ket{LR}_B+ \ket{RL}_A\ket{RL}_B)/\sqrt{2}}, 
\end{equation}
while the other two outcomes leave the system in a separable state. Now, since the measurement of observable $\mathcal{O}_A$ is a local operation it cannot create entanglement, and therefore we can conclude that state~(\ref{eq:doubleavg}) must contain at least an amount of entanglement equal to the average of the entanglement of each of the states that the system can take after measurement,  weighted with the probability of collapsing onto each state. Thus, we have shown that by doubling the two-interferometer device we can recover entanglement between partitions $A$ and $B$ of the extended system, in the presence of noise that is strong enough to destroy the entanglement contained in each of the two devices.

By further doubling the setup, the same principle can be employed to recover entanglement in the case where the noise observed by the different copies of the device varies. We consider first the next order in the expansion of the noise field, for which the phase of the entangled state in each device is not constant but varies linearly with the position of the device. In this case, for a collection of devices that is arranged linearly and spaced evenly, each of the devices would exhibit, for a particular realization of the noise, a state of the form
\begin{equation}
    \Psi_j =  \frac{1}{\sqrt{2}} ( \ket{L}_{Aj} \ket{L}_{Bj} + e^{i [\phi + (j - 1) \Delta \phi]} \ket{R}_{Aj} \ket{R}_{Bj} ).
\end{equation}
 Now, for each pair of consecutive devices we have the combined state
\begin{equation}
\label{eq:coupleGrad}
\Psi_{j,j+1} = \frac{1}{2} \ket{LL}_A \ket{LL}_B + \frac{1}{2} e^{i[2 \phi + (2j -1)\Delta \phi]} \ket{RR}_A \ket{RR}_B + \frac{1}{\sqrt{2}}e^{i [\phi + (j - 1) \Delta \phi]} \frac{\ket{RL}_{A} \ket{RL}_{B} + e^{i \Delta \phi} \ket{LR}_{A} \ket{LR}_{B}}{\sqrt{2}}.
\end{equation}
The second line in Eq.~(\ref{eq:coupleGrad}) shows that the state contains a contribution of a maximally entangled state with a relative phase $\Delta \phi$, which does not depend on the position of the two consecutive devices in the array. Naturally, with increasing size of the fluctuations $\Delta \phi$ the entanglement contained in such a state would be erased upon averaging. However, if we now take two such pairs of devices, that is four devices (8 interferometers), their combined state will contain a component 
\begin{equation}
\Psi_{1-4} = \cdots + \frac{1}{2} e^{i2(\phi + \Delta \phi)}\frac{(\ket{RL}_{A} \ket{RL}_{B} + e^{i \Delta \phi} \ket{LR}_{A} \ket{LR}_{B})_{12}}{\sqrt{2}} \otimes \frac{(\ket{RL}_{A} \ket{RL}_{B} + e^{i \Delta \phi} \ket{LR}_{A} \ket{LR}_{B})_{34}}{\sqrt{2}} + \cdots
\end{equation}
which is of the same form of the state in Eq.~(\ref{eq:pairconst}), and thus we know that it preserves entanglement upon averaging over different realizations of the noise.

The same argument can be applied recursively to cancel higher orders of the noise, where in each iteration the system used for the previous order needs to be doubled, while the recovered entanglement is halved. Notice that in this case the number of copies required to recover in the presence of noise a fraction of the entanglement contained in a single copy is exponential, while in the case described in the main text, aimed at recovering the coherence of a single device, the the scaling was linear. Nevertheless, this example illustrates that the notion of parallelization of interferometers can be applied to the protection of entanglement. Optimizations and further extensions of it will be pursued in future work.

\newpage
\end{widetext}


\begin{thebibliography}{99}

\bibitem{Hasselbach2009} F. Hasselbach, {\it Progress in electron- and ion-interferometry}, \href{https://iopscience.iop.org/article/10.1088/0034-4885/73/1/016101}{Rep. Prog. Phys. {\bf 73}, 016101 (2009)}.

\bibitem{RW2000} H. Rauch and A. Werner, {\it Neutron   Interferometry: Lessons in Experimental Quantum Mechanics (Oxford University, New York)}.

\bibitem{Cornell1998} D. S. Hall, M. R. Matthews, C. E. Wieman, and E. A. Cornell, {\it Measurements of Relative Phase in Two-Component Bose-Einstein Condensates}, \href{https://doi.org/10.1103/PhysRevLett.81.1543}{Phys. Rev. Lett. {\bf 81}, 1543 (1998)}.

\bibitem{Zeilinger1999} M. Arndt, O. Nairz, J. Vos-Andreae, C. Keller, G. van der Zouw, and A. Zeilinger, {\it Wave-particle duality of C$_{60}$ molecules}, \href{https://doi.org/10.1038/44348}{Nature {\bf 401}, 680--682 (1999)}.

\bibitem{Tuxen2013} S. Eibenberger, S. Gerlich, M. Arndt, M. Mayor, and J. T\"uxen, {\it Matter-wave interference of particles selected from a molecular library with masses exceeding 10 000 amu}, \href{https://doi.org/10.1039/C3CP51500A}{Phys. Chem. Chem. Phys. {\bf 15}, 14696--14700 (2013)}.

\bibitem{Arndt2019} Y. Y. Fein, P. Geyer, P. Zwick, F. Kia\l{}ka, S. Pedalino, M. Mayor, S. Gerlich, and M. Arndt, {\it Quantum superposition of molecules beyond 25 kDa}, \href{https://doi.org/10.1038/s41567-019-0663-9}{Nat. Phys. {\bf 15}, 1242-1245 (2019)}.

\bibitem{AspelmeyerKM14} M. Aspelmeyer, T. J. Kippenberg, and C. Marquardt, {\em Cavity Optomechanics}, \href{https://journals.aps.org/rmp/abstract/10.1103/RevModPhys.86.1391}{Rev. Mod. Phys. {\bf 86}, 1391 (2014)}.

\bibitem{TeufelDL2011} J. D. Teufel, T. Donner, D. Li, J. W. Harlow, M. S. Allman, K. Cicak, A. J. Sirois, J. D. Whittaker, K. W. Lehnert, and R. W. Simmonds, {\em Sideband cooling of micromechanical motion to the quantum ground state}, \href{https://doi.org/10.1038/nature10261}{Nature {\bf 475}, 359 (2011)}.

\bibitem{ChanMS2011} J. Chan, T. P. M. Alegre, A. H. Safavi-Naeini, J. T. Hill, A. Krause, S. Gr\"oblacher, M. Aspelmeyer, and O. Painter, {\em Laser cooling of a nanomechanical oscillator into its quantum ground state}, \href{https://doi.org/10.1038/nature10461}{Nature {\bf 478}, 89 (2011)}.

\bibitem{RiedingerWM2018} R. Riedinger, A. Wallucks, I. Marinkovi\'c, C. L\"oschnauer, M. Aspelmeyer, S. Hong, and S. Gr\"oblacher, {\em Remote quantum entanglement between two micromechanical oscillators}, \href{https://doi.org/10.1038/s41586-018-0036-z}{Nature {\bf 556}, 473 (2018)}.

\bibitem{OckeloenDP2018} C. F. Ockeloen-Korppi, E. Damsk\"agg, J.-M. Pirkkalainen, M. Asjad, A. A. Clerk, F. Massel, M. J. Woolley, and M. A. Sillanp\"a\"a, {\em Stabilized entanglement of massive mechanical oscillators}, \href{https://doi.org/10.1038/s41586-018-0038-x}{Nature {\bf 556}, 478 (2018)}.

\bibitem{MillenMP+19} J. Millen, T.S. Monteiro, R. Pettit, and A.N. Vamivakas, {\it Optomechanics with levitated particles}, \href{https://iopscience.iop.org/article/10.1088/1361-6633/ab6100}{Rep. Prog. Phys. {\bf 83}, 026401 (2020)}.

\bibitem{DRD+20} U. Deli\'c, M. Reisenbauer, K. Dare, D. Grass, V. Vuletić, N. Kiesel, and M. Aspelmeyer, {\it Cooling of a levitated nanoparticle to the motional quantum ground state}, 
\href{https://science.sciencemag.org/content/367/6480/892.abstract}{Science {\bf 367}, 892 (2020)}.

\bibitem{Cirac2011} O. Romero-Isart, A. C. Pflanzer, F. Blaser, R. Kaltenbaek, N. Kiesel, M. Aspelmeyer, and J. I. Cirac, {\it Large Quantum Superpositions and Interference of Massive Nanometer-Sized Objects}, \href{https://doi.org/10.1103/PhysRevLett.107.020405}{Phys. Rev. Lett. {\bf 107}, 020405 (2011)}.

\bibitem{Bose2013} M. Scala, M. S. Kim, G. W. Morley, P. F. Barker, and S. Bose, {\it Matter-Wave Interferometry of a Levitated Thermal Nano-Oscillator Induced and Probed by a Spin}, \href{https://doi.org/10.1103/PhysRevLett.111.180403}{Phys. Rev. Lett. {\bf 111}, 180403 (2013)}.

\bibitem{Duan2013} Z.-q. Yin, T. Li, X. Zhang, and L. M. Duan, {\it Large quantum superpositions of a levitated nanodiamond through spin-optomechanical coupling}, \href{https://doi.org/10.1103/PhysRevA.88.033614}{Phys. Rev. A {\bf 88}, 033614 (2013)}.

\bibitem{Albrecht14} A. Albrecht, A. Retzker, and M.B. Plenio, {\it Testing quantum gravity by nanodiamond inter-ferometry with nitrogen-vacancy centers}, \href{https://doi.org/10.1103/PhysRevA.90.033834}{Phys. Rev. A {\bf 90}, 033834 (2014)}. 

\bibitem{Plenio2020} J. S. Pedernales, G. W. Morley, and M. B. Plenio, {\it Motional Dynamical Decoupling for Interferometry with Macroscopic Particles}, \href{https://doi.org/10.1103/PhysRevLett.125.023602}{Phys. Rev. Lett. {\bf 125}, 023602 (2020)}.

\bibitem{Tombesi1997} S. Mancini, V. I. Man'ko, and P. Tombesi, {\it Ponderomotive control of quantum macroscopic coherence},
\href{https://doi.org/10.1103/PhysRevA.55.3042}{Phys. Rev. A {\bf 55}, 3042 (1997)}.

\bibitem{Knight1999} S. Bose, K. Jacobs, and P. L. Knight, {\it Scheme to probe the decoherence of a macroscopic object}, \href{https://doi.org/10.1103/PhysRevA.59.3204}{Phys. Rev. A {\bf 59}, 3204 (1999)}.

\bibitem{Bouwmeester2003} W. Marshall, C. Simon, R. Penrose, and D. Bouwmeester, {\it Towards Quantum Superpositions of a Mirror}, \href{https://doi.org/10.1103/PhysRevLett.91.130401}{Phys. Rev. Lett. {\bf 91}, 130401 (2003)}.

\bibitem{Hornberger2014} M. Arndt and K. Hornberger, {\it Testing the limits of quantum mechanical superpositions}, \href{https://doi.org/10.1038/nphys2863}{Nat. Phys. {\bf 10}, 271--277 (2014)}.

\bibitem{Romero-Isart2011} O. Romero-Isart, {\it Quantum superposition of massive objects and collapse models}, \href{https://doi.org/10.1103/PhysRevA.84.052121}{Phys. Rev. A {\bf 84}, 052121 (2011)}.

\bibitem{Ulbricht2013} A. Bassi, K. Lochan, S. Satin, T. P. Singh, and H. Ulbricht, {\it Models of wave-function collapse, underlying theories, and experimental tests}, \href{https://doi.org/10.1103/RevModPhys.85.471}{Rev. Mod. Phys. {\bf 85}, 471 (2013)}.

\bibitem{Schmoele16} J. Schmöle, M. Dragosits, H. Hepach, M. Aspelmeyer, {\it A micromechanical proof-of-principle 
experiment for measuring the gravitational force of milligram masses}, \href{https://iopscience.iop.org/article/10.1088/0264-9381/33/12/125031/pdf}{Class. Quant. Grav. {\bf 33}, 125031 (2016)}.

\bibitem{Bose2018} S. Qvarfort, A. Serafini, P. F. Barker, and S. Bose, {\it Gravimetry through non-linear optomechanics}, \href{https://doi.org/10.1038/s41467-018-06037-z}{Nat. Commun. {\bf 9}, 3690 (2018)}.

\bibitem{Li2020} M. Rademacher, J. Millen,  and Y. L. Li, {\it Quantum sensing with nanoparticles for gravimetry: When bigger is better}, \href{https://doi.org/10.1515/aot-2020-0019}{Advanced Optical Technologies {\bf 9}, 227–239 (2020)}.

\bibitem{WestphalHP+21} T. Westphal, H. Hepach, J. Pfaff, and M. Aspelmeyer, {\it Measurement of gravitational coupling between millimetre-sized masses}, \href{https://www.nature.com/articles/s41586-021-03250-7}{Nature {\bf 591}, 225 (2021)}.

\bibitem{Plenio2020+} J. S. Pedernales, F. Cosco, and M. B. Plenio, {\it Decoherence-Free Rotational Degrees of Freedom for Quantum Applications}, \href{https://doi.org/10.1103/PhysRevLett.125.090501}{Phys. Rev. Lett. {\bf 125}, 090501 (2020)}. 

\bibitem{Plenio2021} F. Cosco, J. S. Pedernales, and M. B. Plenio, {\it Enhanced force sensitivity and entanglement in periodically driven optomechanics}, \href{https://doi.org/10.1103/PhysRevA.103.L061501}{Phys. Rev. A {\bf 103}, L061501 (2021)}.

\bibitem{Plenio2021+} J. S. Pedernales, K. Streltsov, and M. B. Plenio, {\it Enhancing Gravitational Interaction between Quantum Systems by a Massive Mediator}, \href{https://arxiv.org/abs/2104.14524}{arXiv:2104.14524 (2021)}.

\bibitem{PMP20} J. S. Pedernales, G. W. Morley, and M. B. Plenio, {\it Motional Dynamical Decoupling for Matter-Wave Interferometry}, \href{https://arxiv.org/abs/1906.00835}{arXiv:1906.00835 (2019)}.

\bibitem{PPS+18} H. Pino, J. Prat-Camps, K. Sinha, B. P. Venkatesh, and O. Romero-Isart, {\it On-chip quantum interference of a superconducting microsphere}, \href{https://iopscience.iop.org/article/10.1088/2058-9565/aa9d15}{Quantum Sci. Technol. {\bf 3}, 25001 (2018)}.

\bibitem{SuppMat} See Supplemental Material for a detailed derivation of the probability distribution and extensions to the case of entangled interferometers.

\bibitem{Plenio2005} M. B. Plenio, {\it The logarithmic negativity: A full entanglement monotone that is not convex}, \href{https://doi.org/10.1103/PhysRevLett.95.090503}{Phys. Rev. Lett. {\bf 95}, 090503 (2005)}


\end{thebibliography}
\end{document}